# Detecting new obfuscated malware variants: A lightweight and interpretable machine learning approach

**Authors:** Oladipo A. Madamidola[a], Felix Ngobigha[a] and Adnane Ez-zizi[a,*]

**Affiliations:** [a] University of Suffolk, Waterfront Building, IP4 1QJ, Ipswich, UK

## Abstract

Machine learning has been successfully applied in developing malware detection systems, with a primary focus on accuracy, and increasing attention to reducing computational overhead and improving model interpretability. However, an important question remains underexplored: How well can machine learning-based models detect entirely new forms of malware not present in the training data? In this study, we present a machine learning-based system for detecting obfuscated malware that is not only highly accurate, lightweight and interpretable, but also capable of successfully adapting to new types of malware attacks. Our system is capable of detecting 15 malware subtypes despite being exclusively trained on one malware subtype, namely the Transponder from the Spyware family. This system was built after training 15 distinct random forest-based models, each on a different malware subtype from the CIC-MalMem-2022 dataset. These models were evaluated against the entire range of malware subtypes, including all unseen malware subtypes. To maintain the system's streamlined nature, training was confined to the top five most important features, which also enhanced interpretability. The Transponder-focused model exhibited high accuracy, exceeding 99.8%, with an average processing speed of 5.7 µs per file. We also illustrate how the Shapley additive explanations technique can facilitate the interpretation of the model predictions. Our research contributes to advancing malware detection methodologies, pioneering the feasibility of detecting obfuscated malware by exclusively training a model on a single or a few carefully selected malware subtype and applying it to detect unseen subtypes.

* Corresponding author. Email address: a.ez-zizi@uos.ac.uk (A. Ez-zizi)

# 1. Introduction

Over the last two decades, technological advancements in cloud computing, the Internet of Things (IoT) and the introduction of fifth Generation (5G) and beyond 5G mobile networks have revolutionised the way businesses and individuals access and store data (Mijwil et al., 2023). This technological paradigm shift has enabled individuals and organisations to access their data seamlessly from anywhere in the world, using any connected devices. However, malware (i.e., malicious software) poses a significant threat to the security of these technologies. Malicious actors can use malware to compromise the confidentiality, integrity, and availability of data (Gupta and Rani, 2020). The impact of malware can be devastating for businesses and individuals alike, as it can result in the loss of sensitive information, such as personal data and financial information. In 2021 alone, over 1.3 billion malware specimens were detected (Dener et al., 2022), and with increased connectivity, reliance on digital systems, and the growing number of connected devices, the attack landscape is expected to grow even more.

With an unprecedented number of malware targeting various computing systems and online infrastructures, the detection of malware is of great importance. Unfortunately, traditional methods of malware detection, such as signature-based detection and behaviour-based detection, are becoming increasingly less effective against modern and sophisticated malware attacks. Malicious authors are using advanced technologies to design malware that is increasingly difficult to detect and exterminate (Mezina and Burget, 2022). In recent years, there has been a pivot towards the use of machine learning (ML) for malware detection. One of the advantages of the ML-based approach is its capacity to process a large number of files quickly to identify patterns and anomalies that may be indicative of malware attacks. The classical ML-based approach to malware detection typically involves training a model on a dataset of known malware and benign software and then using the model to classify new samples from the same distribution of malware and benign types (see, for example, Alani et al., 2023; Shafin et al., 2023; Roy et al., 2023). While these techniques have been shown to learn from past attacks effectively, the following question remains underexplored: Are these ML-based models capable of detecting entirely new types of malware not represented in the training data?

## 1.1. Research contributions

In their 2023 study, Alani et al. underscored the essential pillars for effective malware detection systems: (1) high accuracy, (2) lightweight design (i.e. minimal memory and processing resource requirements) and (3) explainability (as opposed to opaque, 'black box' approaches). Here, we contend that a fourth pillar—adaptability to novel malware threats—is equally crucial in the face of the continuously evolving cyber-threat landscape. We thus designed a malware detection system incorporating all four aforementioned criteria. Pioneering a zero-shot machine-learning approach (Palatucci et al., 2009) in malware detection, our system was trained and tested on the CIC-MalMem-2022 dataset, with a focus on assessing its capacity to detect unseen obfuscated malware variants. Our primary contributions include:

1. Development of a comprehensive and adaptive machine learning-based system for detecting obfuscated malware, with high accuracy (> 99.8%) and a fast detection rate (5.7 μs per file), while trained on a small fraction of the dataset (<7%) used in other related works.
2. Use of a novel training methodology, where the system, initially trained on a single malware subtype, can successfully detect 14 distinct, previously unencountered obfuscated malware subtypes. This approach demonstrates the ability of machine learning-based systems in identifying and combating zero-day malware threats.
3. Implementation of feature selection, leading to a reduced memory usage (a model size of 340 KB) and a simplification of the detection system's complexity without compromising accuracy.
4. Inclusion of global and local interpretations of the system's predictions to demonstrate how practitioners can gain deeper insights. This should encourage a more transparent and informed approach to automated cyber security defences.

## 1.2. Structure of paper

The remaining sections of this paper are as follows. Section 2 presents a brief review of relevant literature and background content on malware detection. In Section 3, we provide details of the study methodologies. Section 4 covers the performance results of the various models considered. Finally, Section 5 provides a comprehensive discussion and comparative analysis of the results of this study, as well as limitations and potential future directions.

# 2. Literature review and background

## 2.1. Malware types

Malware is a type of software designed to harm or exploit any device, network or system it infects. Malicious actors use them to gain unauthorised access to steal sensitive information (e.g., financial data), cause service disruptions or establish remote control access for future exploitation, among other types of harm. Although malware manifests in diverse forms, it typically falls into a known set of distinct types, each with unique characteristics and objectives. These malware types can overlap, with many types showing traits of other malware types. The most prevalent types include Trojan horse, Spyware and Ransomware.

**Trojan Horse:** Trojan horses appear and behave like legitimate software to deceive users into executing them. Once activated, they can carry out malicious activities in the background, including stealing sensitive information through keylogging, monitoring user activities and altering files within the systems it resides in (Idika and Mathur, 2007). Trojan horses are generally propagated via the download of apps people consider legitimate. The dataset used in this study includes the following Trojan horse subtypes: Zeus, Emonet, Refroso, Scar and Reconyc.

**Spyware:** This is a class of malware used to secretly record user activities or steal personal information (such as browsing habits or activities) that can be sold to third parties (e.g., for custom advertising; Wang et al., 2006). 180Solutions, CoolWebSearch, Gator, Transponder and TIBS are subtypes of spyware included in the dataset used in this study.

**Ransomware:** This is a class of malware that is used to take control of a computer by encrypting all data on a computer system (Tahir, 2018). As a result of encryption, the user cannot access their data or use their operating system. The screen of an infected computer is usually used by an attacker to make their demands such as the payment of money (though the victim's response to the ransom demands does not guarantee recovery). The dataset used in this study includes Conti, Maze, Pysa, Ako and Shade as subtypes of ransomware.

## 2.2. Malware obfuscation techniques

While malware detection poses a significant challenge, malware authors have recently exacerbated this challenge by employing obfuscation techniques to make their code more intricate and resistant to detection (O'Kane et al., 2011). Obfuscation can take many forms, such as *code obfuscation*, which involves making the code difficult to understand and analyse, for example, by renaming variables and functions, adding unnecessary code or using complex control structures (Rad et al., 2012); *code encryption*, which encrypts the malware code using a secret key or algorithm to evade detection by antivirus software, as the encrypted code may not match known malware signatures (Rad et al., 2012); *polymorphism*, where the malware is designed to change its code structure and behaviour with each new infection (Alam et al., 2015); and *metamorphism*, which goes a step further than polymorphism by changing its code structure and behaviour even while it is running (O'Kane et al., 2011).

## 2.3. Malware analysis approaches

In the field of cyber security, malware analysis is primarily performed using two methodologies: static analysis and dynamic analysis, each with its distinct mechanisms and implications (Aslan and Samet, 2020; Elkhail et al., 2021). Both static and dynamic analysis approaches result in the generation of excessively large numbers of features and signatures. To mitigate this complexity and battle the unprecedented increase in the number of malware specimens, researchers in recent years have started using ML techniques to improve malware detection based on features generated from static and dynamic analysis (Dada et al., 2019). The remaining subsections will delve into some notable studies that have leveraged ML for malware analysis, with a focus on binary classification of benign versus malware software and the detection of zero-day attacks.

### 2.3.1 Machine learning-based static analysis

Using a dataset derived from Windows Portable Executable files (PE files), Liu et al. (2020) proposed a malware detection system based on adversarial training, which achieved up to 97.73% accuracy. Similarly, Radwan (2019) performed malware detection using data from the static analysis of PE files. The dataset consisted of 2683 malware, 2501 benign records and 55 variables. After training seven classifiers—Gradient Boosted Trees, Decision Tree, Random Forest (RF), K-Nearest Neighbours, File large margin, Logistic Regression and Naïve Bayes—the authors showed that RF technique performed the best, with a detection accuracy of 99.23%.

In an attempt to detect harmful mobile applications, Huang et al. (2013) evaluated the performance of AdaBoost, Naïve Bayes, Decision Tree and Support Vector Machine for malware classification based on data generated from the permission calls of an application. Their results suggest that the Naïve Bayes technique can be used to detect more than 81% of malicious samples (Huang et al., 2013). Although signature-based approaches are effective against known malware signatures, they can be ineffective in detecting zero-day malware that lacks pre-existing signatures (Aslan and Samet, 2020; Elkhail et al., 2021). In addition, recent malware now uses obfuscation techniques, such as modifying code, replacing instructions, reassigning registers and inserting redundant code, all of which are aimed at evading detection. Therefore, this strategy can be ineffective in identifying malicious software that uses obfuscation, packaging or encryption methods (Sihwail et al., 2019).

### 2.3.2 Machine learning-based dynamic analysis

In their dynamic analysis, Bhatia and Kaushal (2017) employed the Decision Tree and RF algorithms to classify datasets derived from system call traces of Android applications as malicious or benign. Their results indicated that the Decision Tree and RF algorithms achieved accuracies of 85% and 88% respectively. In a subsequent study, Hwang et al. (2020) developed a two-stage detection model. First, they use an analytical framework, the Markov model, to dynamically characterise Windows API calls of ransomware. Next, they applied ML techniques to classify the data. The authors demonstrated that they were able to achieve an overall accuracy of 97.3% with 4.8% false positives and 1.5% false negatives using a RF algorithm. While dynamic-based approaches offer a deeper understanding of the true nature of the malware and the threat it poses without the risk of infection of the entire enterprise architecture. However, due to advancements in technology, malware is becoming more complex, and evasion techniques are being adopted by adversaries to evade detection in sandboxes by ensuring that malware remains dormant until a certain trigger is activated (Sihwail et al., 2019).

### 2.3.3 Machine learning-based hybrid analysis

To address the issue of dormant malware, other studies have used features from hybrid analysis, that is, a combination of features from dynamic analysis and static analysis with ML techniques. For example, Ijaz et al. (2019) extracted 2300 features from dynamic analysis, and 92 features from the static analysis of PE files. Owing to the large number of features, different combinations of features were evaluated using ML algorithms. They reported that the Gradient Boosting Algorithm achieved 94.64% accuracy on data from the dynamic analysis, whereas the accuracy of the static analysis-based model was 99.36%. Hadiprakoso et al. (2020) also reported that the Gradient Boosting Algorithm is effective for classifying malware that targets Android applications. Using their method, a detection accuracy of approximately 99% was achieved using a Gradient Boosting Algorithm.

### 2.3.4 Machine learning-based memory analysis

Owing to the increasingly changing behaviour of malware, data from forensic memory analysis has been proposed for effective malware detection in recent years (Mosli et al., 2016, Rathnayaka and Jamdagni, 2017, Sihwail et al., 2021). Memory-based features have been suggested as an effective way to overcome some of the limitations of other methods of malware detection (Dener et al., 2022). Useful information such as active and terminated processes, Dynamic Link Libraries (DLL) used, running services, registry entries, and active network connections can readily be read from memory (Rathnayaka and Jamdagni, 2017; Sihwail et al., 2019; Sihwail et al., 2021). Furthermore, memory analysis can help identify attackers' IP addresses, hooks used to hide themselves, malware injections, and interdependencies of processes (Rathnayaka and Jamdagni, 2017)

In 2022, Carrier et al. proposed a stacked ensemble system to classify obfuscated malware by using features derived from device memory of recent and advanced obfuscated malware attacks. In their ensemble system, they use Naïve-Bayes, RF and Decision Tree as base learners and Logistic Regression as meta-learner. With this approach, they reported 99% accuracy for malware detection. Their work also resulted in the publication of a new dataset specifically designed to test the detection of obfuscated malware, namely the CIC-MalMem-2022 dataset. Using the same dataset, Dener et al. (2022) proposed a detection method within a big data environment. In their study, they evaluated machine and deep learning techniques for binary malware detection. Results were evaluated based on accuracy, F1-score, precision, recall, and AUC performance metrics. The authors reported that the Logistic Regression algorithm achieved 99.97% accuracy, and the Gradient Boost Tree achieved 99.94% accuracy for malware detection by memory analysis (Dener et al., 2022).

Improving on the work by Carrier et al. (2022), Alani et al. (2023) proposed a ML-based system that uses the recursive feature elimination method to reduce the number of features initially proposed by Carrier et al. (2022) from 55 to 5. The selected features were then used to train an Extreme Gradient Boost classifier, with an accuracy of over 99% and a speed of malware detection of 0.413 μs. Another contribution of Alani et al.'s study is the integration of an AI explainability approach based on Shapely additive explanations to support the interpretation of model predictions.

Although identifying malware is crucial, there is an argument that discerning the specific type of malware can enhance the responsiveness and effectiveness of anti-malware systems, while also influencing the approach taken to mitigate the malware attack. Using deep learning techniques, Mezina and Burget (2022) develop a dilated convolutional neural network for binary and multiclass classification (Benign and Malware types) and detection of obfuscated malware using the CIC-MalMem-2022 dataset. They found that RF outperforms the dilated convolutional neural network model for binary classification (with 99% accuracy). Similarly, Roy et al. (2023) and Shafin et al. (2023) propose different systems for detecting obfuscated malware types and malware subtypes using the CIC-MalMem-2022 dataset. With their MalHyStack method, Roy et al. (2023) reported an accuracy of 99.97% for binary classification. On the other hand, Shafin et al. (2023) reported 99.96% accuracy for binary classification by their RobustCBL method and 99.92% accuracy by their CompactCBL method respectively.

### 2.3.5 Machine learning-based detection of zero-day attacks

Detection of previously unseen malware types, or what is often referred to as zero-day malware, represents a significant and growing challenge within the cyber security field. It is estimated that approximately 350,000 instances of zero-day malware are generated daily (Amer and Zelinka, 2020). These zero-day malware attacks exploit unknown system vulnerabilities and uses evasion techniques to evade cyber security detection tools. Several ML-based solutions have been proposed (for a recent review, see Guo, 2023), ranging from supervised learning (Alazab et al., 2011, Zhou and Pezaros, 2019, Gandotra et al., 2016), unsupervised outlier detection approaches (Kim et al., 2018; Mirsky et al., 2018) and semi-supervised learning approaches (Huda et al., 2017) to reinforcement learning (Acuto et al., 2023).

Jain and Singh (2017) proposed an integrated approach that use features from both static and dynamic analysis for detection and classification of zero-day malware. They evaluated three classifiers (Naïve Bayes, RF and Support Vector Machine), and observed that RF achieved the best accuracy of 73.47% for the detection of zero-day malware (Jain and Singh, 2017). Using the CSE-CIC-IDS 2018 dataset, Zhou and Pezaros (2019) evaluate the effectiveness of RF, Naïve Bayes, Decision Tree, Multi-layer Perceptron, K-Nearest Neighbours and Quadratic Discriminant for the detection of 14 different intrusions attacks. To stimulate zero-day attacks, the trained classifiers were tested on eight new attacks that were not included in training but were collected from real-world attack scenarios. An evaluation of various classification algorithms revealed performance discrepancies. The Decision Tree model emerged as the best performer, achieving 96% accuracy on unseen attacks. Alhaidari et al. (2022) proposed a zero-day vigilante system (ZeVigilante) for the detection of unknown malware. In their study, they considered both static and dynamic analysis and evaluated the performance of RF, Neural Networks, Decision Tree, K-Nearest Neighbours, Naïve-Bayes and Support Vector Machine for the detection of unknown malware. They reported that RF achieved the highest accuracy of 98.17% and 98.89% for static and dynamic analysis respectively.

Although the proposed zero-day attack detection systems mentioned above showed promise, they often exhibit large variation in detection accuracy against different types of attacks, and they lack explainability. Also, apart from Alazab et al (2011), these studies have not considered issues of obfuscation, which is one of the major evasion techniques used by newer malware specimens. Additionally, none of the previous studies have used memory-analysis based features or considered interpretable ML approaches. Table 1 provides a concise summary of related work that was reviewed and compared in this study.

*Table 1. Comparison and summary of related work. * For the study by Alazab et al. (2011), we present the weighted F1-score since accuracy was not reported in their paper.*

| Approach | Research | Classifier | Accuracy | Obfuscation | Interpretability | Zero Day Attacks |
|---|---|---|---|---|---|---|
| Static Analysis | Liu et al. (2020) | Visual-AT | 0.9773 | x | x | x |
| | Radwan (2019) | Random Forest | 0.9923 | x | x | x |
| | Huang et al. (2013) | Naïve Bayes | 0.8100 | x | x | x |
| | Alazab et al. (2011) | Support Vector Machine | 0.9840* | ✓ | x | ✓ |
| | Zhou and Pezaros (2019) | Decision Tree | 0.9600 | x | x | ✓ |
| | Alhaidari et al. (2022) | Random Forest | 0.9817 | x | x | ✓ |
| Dynamic Analysis | Bhatia and Kaushal (2017) | Decision Tree | 0.8500 | x | x | x |
| | | Random Forest | 0.8800 | | | |
| | Hwang et al. (2020) | Random Forest | 0.9730 | x | x | x |
| | Alhaidari et al. (2022) | Random Forest | 0.9889 | x | x | ✓ |
| Hybrid Analysis | Ijaz et al. (2019) | Gradient Boosting | 0.9464 | x | x | x |
| | Hadiprakoso et al. (2020) | Gradient Boosting | 0.9900 | x | x | x |
| | Gandotra et al. (2016) | Random Forest | 0.9997 | x | x | ✓ |
| | Jain and Singh (2017) | Random Forest | 0.7347 | x | x | ✓ |
| Memory Analysis | Carrier et al. (2022) | Ensemble | 0.9900 | ✓ | x | x |
| | Dener et al. (2022) | Logistic Regression | 0.9997 | ✓ | x | x |
| | Mezina and Burget (2022) | Dilated Convolutional Neural Network | 0.9989 | ✓ | x | x |
| | Shafin et al. (2023) | RobustCBL | 0.9996 | ✓ | x | x |
| | | CompactCBL | 0.9992 | | | |
| | Roy et al. (2023) | Hybrid Stack Ensemble | 0.9998 | ✓ | x | x |
| | Alani et al. (2023) | Extreme Gradient Boost | 0.9985 | ✓ | ✓ | x |

## 2.4. Summary

While ML techniques have shown promise in recognising unique malware signatures and classifying known malware samples, there is a need to extend these methods to effectively detect previously unseen obfuscated malware types, particularly when using data from memory analysis. Furthermore, it is essential to establish a comprehensive framework for developing malware detection systems. Alani et al. (2023) proposed a framework based on three critical criteria: (1) high accuracy, (2) lightweight design (i.e. minimal memory and processing resource requirements) and (3) explainability (as opposed to opaque, 'black box' approaches). Here, we argue that a fourth criteria—adaptability to novel malware threats—is indispensable, given the continuously evolving cyber-threat landscape. We thus designed a malware detection system incorporating all four aforementioned criteria. Our system was trained and tested on the CIC-MalMem-2022 dataset, with a focus on assessing its capacity to detect unseen obfuscated malware.

# 3. Methodology

## 3.1. Dataset

The CIC-Malmem-2022 dataset used in this study is a publicly available dataset introduced by Carrier et al. (2022). The dataset was created from memory dumps of recent real-life cyberattacks. It has a total of 58,596 records, evenly split between 29,298 benign and 29,298 malicious instances. Each instance consists of 55 features extracted from the single memory dump file using VolMemLyzer (Ahlashkari, 2022). Features include, for example, the number of running processes, the number of open dynamic-link libraries (DLLs), the average number of threads per process and the number of open files (for more detailed information about the features, refer to Table A.1 in Appendix). The dataset further classifies each malicious file by malware type (Ransomware, Spyware or Trojan Horse) and subtype (e.g., Zeus, Gator, Pysa, etc). Table 2 summarises the distribution of malware types and subtypes in the dataset.

*Table 2. Summary* of malware types and subtypes distribution in the CIC-Malmem-2022 dataset.

| Malware Type | Malware Subtype | Number of Instance | Percentage (%) |
|---|---|---|---|
| Trojan Horse | Zeus | 1950 | 3.3 |
| | Emotet | 1967 | 3.4 |
| | Refroso | 2000 | 3.4 |
| | Scar | 2000 | 3.4 |
| | Reconyc | 1570 | 2.7 |
| Spyware | 180Sulotions | 2000 | 3.4 |
| | CoolWebSearch | 2000 | 3.4 |
| | Gator | 2200 | 3.8 |
| | Transponder | 2410 | 4.1 |
| | TIBS | 1410 | 2.4 |
| Ransomware | Conti | 1988 | 3.4 |
| | MAZE | 1958 | 3.3 |
| | Pysa | 1717 | 2.9 |
| | Ako | 2000 | 3.4 |
| | Shade | 2128 | 3.6 |
| **Total** | - | **29298** | **50.0** |

## 3.2. Data modelling and evaluation metrics

### 3.2.1. Data pre-processing

Standard procedures were employed to prepare the dataset for ML modelling, including the removal of invariant features (*pslist.nprocs64bit*, *handles.nport* and *svcscan.interactive_process_services*), scaling numerical features using the min-max method and label-encoding the categorical target variable (0 for benign and 1 for malware). The dataset did not contain any missing values.

### 3.2.2. Evaluation criteria

The performance of the classifiers was examined using the following metrics: confusion matrix, accuracy, precision, recall and F-1 score. These performance metrics are defined below.

**Confusion matrix:** A matrix that summarises the performance of a ML model on a set of test data. It provides a graphical display of the number of accurate and inaccurate model's predictions, broken down by class. In our binary classification problem, the confusion matrix is divided into four quadrants displaying the following metrics: True Positive (TP), True Negative (TN), False Positive (FP), and False Negative (FN).

**Accuracy:** The ratio of correctly classified instances divided by the total number of instances. It is measured using the following equation:

$$Accuracy = \frac{(TP+TN)}{(TP+FP+TN+FN)} \quad (1)$$

**Precision:** The ratio of true positives divided by true positives and false positives:

$$Precision = \frac{TP}{TP+FP} \quad (2)$$

**Recall (also known as sensitivity):** The ratio of true positives divided by true positives and false negatives:

$$Recall = \frac{TP}{TP+FN} \quad (3)$$

**F1-Score:** The harmonic means of precision and recall:

$$F1\text{-}Score = 2 \times \frac{Precision \times Recall}{Precision + Recall} \quad (4)$$

### 3.2.3. Random Forest and feature selection

RF, an ensemble method that can be used for both classification and feature selection, is the main ML algorithm used in this study, following an earlier assessment of various ML algorithms. It works by constructing multiple decision trees during training and aggregating their predictions (Genuer et al., 2010). Each tree is trained on a bootstrap sample, and optimal features at each split are identified from a random subset of all features (Degenhardt et al., 2019). The final features are determined from all trees based on the mean decrease in accuracy value obtained from multiple calculations (Zhao et al., 2022). The feature importance derived from RF—an embedded selection method (Speiser et al., 2019, Alduailij et al., 2022)—was particularly beneficial for our lightweight system, allowing the use of an optimal subset of features without compromising on accuracy. The advantage of using a RF for feature selection is that it is fast to train and is robust to noise (Niu et al., 2020).

## 3.3. Experimental methodology

The development process of our malware detection system can be divided into the following three stages as described below (see Figure 1 for a summary).

### 3.3.1. Stage 1: Development of a baseline classifier

Here, baseline performance results are obtained for the task of classifying benign versus malware. To achieve this, the original dataset is split into training (80%) and test (20%) sets. The original representations of the malware subtypes in the training and test samples are retained using stratification. Seven classifiers are trained and tested on the processed dataset. The best-performing classifier is determined based on the F1 score. To validate the results, a stratified 10-fold cross-validation method is also used. All in all, this stage represents the traditional way of developing ML-based malware detection systems, where the model is trained and tested on the same malware (sub-)types.

### 3.3.2. Stage 2: Development of an adaptive malware detection system

Once a baseline model for classifying malware and benign memory dumps has been established, we turn to the objective of building an adaptive malware detection system that can detect new malware attacks. Furthermore, we want our system to also be highly accurate, interpretable and lightweight (in the sense that is fast and relies on a small number of features). To achieve this, 15 distinct models are built, each on a different malware subtype. The training data contains only 80% of malware instances from a specific malware subtype (e.g., 80% of Transponder instances) along with an equivalent randomly selected number of benign instances. The trained model is then tested on a holdout unseen dataset containing the remaining benign instances and all other malware subtypes in addition to the remaining 20% of the specific malware instances. To make the system lightweight, the top five features selected based on RF-based feature importance are retained for the training and testing processes.

### 3.3.3. Stage 3: Model interpretation

Here we demonstrate how the model's decision-making process can be interpreted using the Shapely Additive Explanations technique both at a global (i.e. how to interpret the effect of the main features) and a local level (i.e. how to interpret a single prediction).

## 3.4. Apparatus

The experimentation setup for this study included the following hardware specifications: an Intel(R) Core (TM) i5-1035G1 CPU @ 1.00GHz 1.19 GHz processor, 8.00 GB of RAM, Intel(R) UHD Graphics GPU, Windows 10 Home operating system. For software implementation, Python v3.11.5 was used with the following packages: Pandas (v1.3.5) for data manipulation and analysis, Scikit-learn (v1.2.2) for running ML algorithms, Matplotlib (v3.6.2) for plotting graphs, and SHAP (v0.42.1) for model explainability.

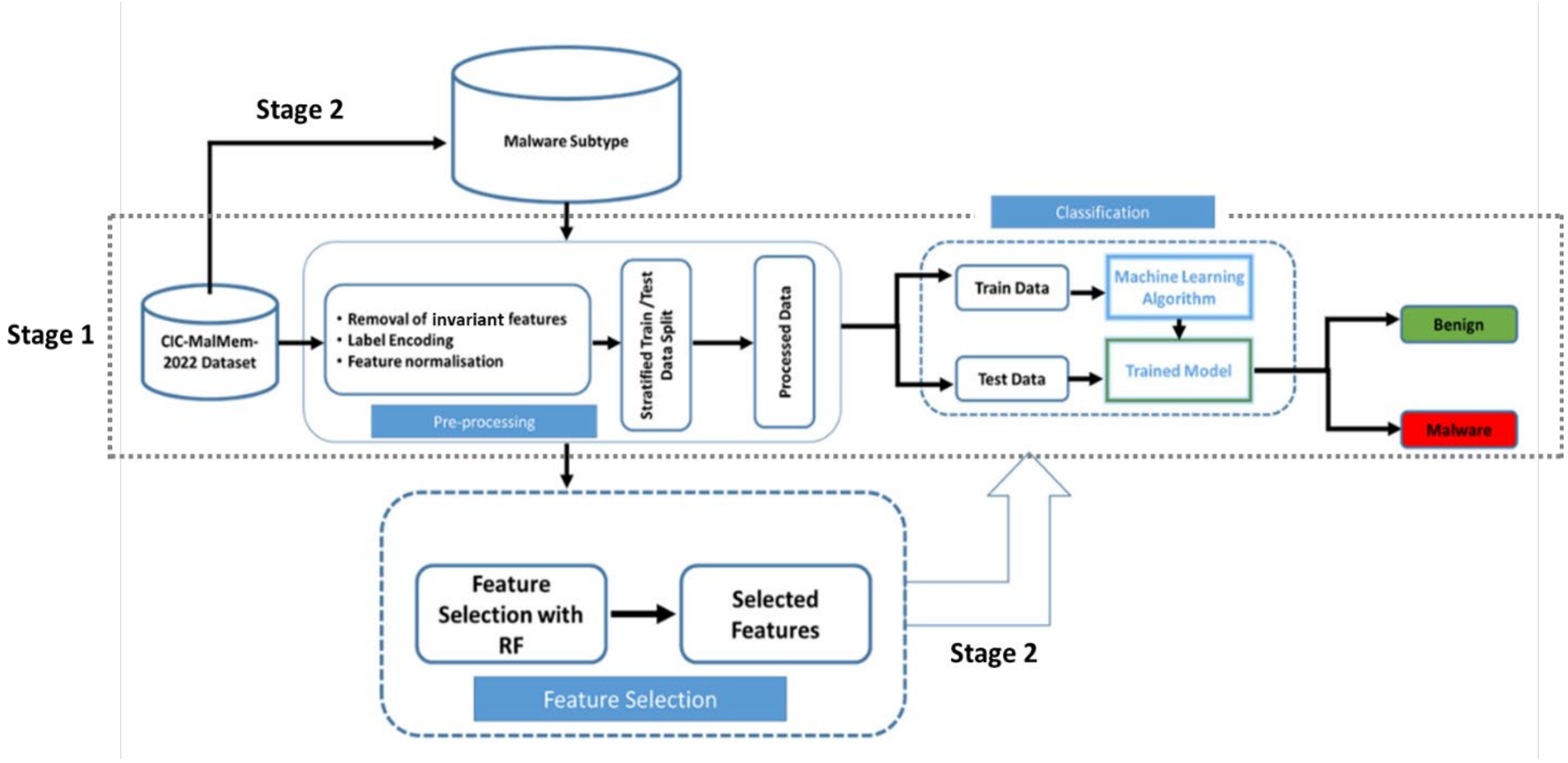

*Figure 1. Overview of the development process of the proposed obfuscated malware detection system. The process begins with the development of a baseline classifier using all malware data (Stage 1). The next phase (Stage 2) consists of developing a separate classifier for each malware subtype, using only the top five features selected based on feature importance.*

# 4. Results

## 4.1. Assessing baseline classifier performance against known malware variants

As explained in Section 3.3.1, in the first stage, seven classifiers— RF, Gradient Boosting, Decision Tree, K-Nearest Neighbours, Logistic Regression, Support Vector Machine and Naive Bayes—were evaluated for binary malware detection. After removing constant features (*pslist.nprocs64bit*, *handles.nport* and *svcscan.interactive_process_services*), the remaining 52-feature dataset was randomly split into 80% training and 20% testing subsets. All subtypes of malware were included in the training dataset and subsequently tested, as in previous studies (see, for example, Alani et al., 2023; Dener et al., 2022; Mezina and Burget, 2022; and Shafin et al., 2023). The performance metrics, as detailed in Table 3, revealed that all seven classifiers were effective at detecting malware, with average accuracies surpassing 99.2%. Notably, RF achieved marginally higher accuracy compared to the other classifiers, and thus was selected as the best-performing classifier and subsequent experiments were based on it.

*Table 3. Binary classification results for models tested on previously encountered malware subtypes. Findings from previous studies are listed at the bottom of the table for comparative purposes.*

| Classifiers | Accuracy | Precision | Recall | F1 Score |
|---|---|---|---|---|
| Random Forest | 1.0000 | 1.0000 | 1.0000 | 1.0000 |
| Gradient Boosting | 0.9996 | 0.9996 | 0.9996 | 0.9996 |
| Decision Tree | 0.9997 | 0.9997 | 0.9997 | 0.9997 |
| Support Vector | 0.9970 | 0.9970 | 0.9970 | 0.9970 |
| Logistic Regression | 0.9958 | 0.9958 | 0.9958 | 0.9958 |
| GuassianNB | 0.9922 | 0.9922 | 0.9922 | 0.9922 |
| Kneighbors | 0.9998 | 0.9998 | 0.9998 | 0.9998 |

To validate our results, we also run 10-fold cross-validation, which produced comparable results with all F1-scores above 0.999 as shown in Table A.2 in Appendix. This demonstrates the robustness and reliability of the developed models for classifying known malware variants. Next, we assess the capability of such ML models to detect unseen malware variants.

## 4.2. Performance analysis of the adaptive malware detection system

Having established the baseline performance results, we aimed to develop a malware detection system capable of detecting novel malware (sub-)types, while maintaining high accuracy and lightweight design architecture. Here, we sought to adopt a strategy that is radically different from what has been proposed previously in the literature. As explained in Section 3.3.2, the objective in this second stage was to assess if models trained on a reduced dataset from one malware subtype can be used to identify emerging threats (i.e., be future-proof). We postulated that each malware subtype will exhibit distinct obfuscation traits, therefore leaving different signatures in memory, and as a result, will have different features' importance rankings.

To test this, 15 new models were trained, each on 80% of data from a specific malware subtype (e.g., 80% of Transponder instances), along with an equivalent randomly selected number of benign instances, as explained in Section 3.3.2. Following the approach of Alani et al. (2023) to make the models lightweight, only the top five features from the RF feature importance were used in the training and testing of each model (also our experiments show that increasing the number of features above five only marginally affect the model performance, as can be seen in Figure A.1). The results are presented in Figure 2 and Table 4 (for more detailed information about feature ranking of each malware subtype-based model, refer to Table A.2 in Appendix).

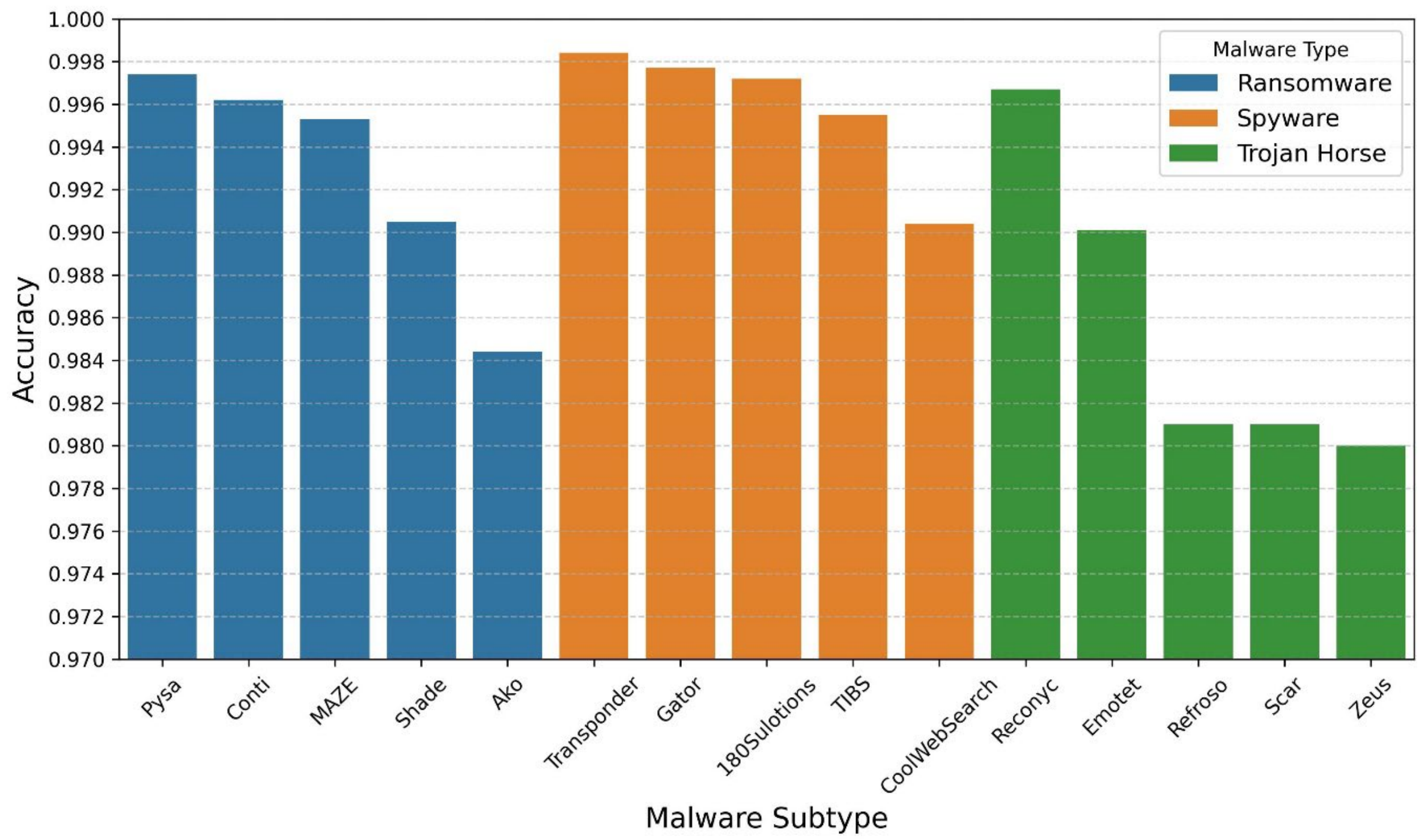

*Figure 2. Comparison of accuracies of the different malware subtype-based models.*

The model achieved an accuracy of over 99% for 11 out of 15 malware subtypes, with the Transponder malware subtype producing the highest accuracy at 99.84%, and the lowest accuracy for Zeus at 98% (Figure 2). Looking at the selected features (Table 4), *svcscan.nservices* and *svcscan.shared_process_services* are in the top three features for most of the malware subtypes, and they are the top two most important features for six malware subtype-based models (namely, Emotet, Reconyc, Gator, TIBS, Pysa and Shade; see Table A2). Also, as shown in Table 4, only 10 features out of 52—representing four feature categories—were selected in the 15 malware subtype-focused models. This indicates that malware attacks, albeit belonging to their different variants, tend to target common memory processes and share common characteristics.

*Table 4. Overall frequencies of selected features from the 15-malware subtype-focused models, along with the frequency breakdown by malware type.*

| Feature Category | Feature | Description | Feature Importance Ranking | | |
|---|---|---|---|---|---|
| | | | 1st | 2nd | 3rd |
| SVCscan | svcscan.**nservices** | Total number of services running. | 10 | 1 | 4 |
| SVCscan | svcscan.**shared_process_services** | Total number of services in shared processes. | 0 | 7 | 4 |
| SVCscan | svcscan.**kernel_drivers** | Total number of kernel drivers. | 1 | 0 | 3 |
| PsList | pslist.**avg_handlers** | Average number of handlers. | 2 | 3 | 1 |
| Handles | handles.**avg_handles_per_proc** | Average number of handles per process. | 1 | 3 | 1 |
| Handles | handles.**nevent** | Total number of event handles. | 0 | 1 | 0 |
| Handles | handles.**nmutant** | Total number of mutant handles. | 0 | 0 | 0 |
| Handles | handles.**nhandles** | Total number of handles opened. | 1 | 0 | 0 |
| DLLlist | dlllist.**avg_dlls_per_proc** | Average number of DLLs loaded per process. | 0 | 0 | 0 |
| Handles | handles.**nsection** | Total number of section handles. | 0 | 0 | 1 |

The Transponder malware subtype, which represents the malware subtype producing the highest accuracy, also exhibited a high speed in classifying memory dump instances, with an average speed of 5.7 µs per instance. This speed is comparable to that reported by Alani et al. (2023) for their RF model, which had a processing speed of 5.2 µs per memory dump. Furthermore, the model size of 340 KB demonstrates the feasibility of deploying such a system in resource-constrained devices.

Figure 3 depicts the resulting confusion matrix, which indicates that the false-negative rate was only 0.01% (i.e., 8 instances were wrongly classified as benign out of 27370 unseen malware instances) and that the false-positive rate was 0.15% (i.e., 81 instances out of the 27370 benign instances were wrongly classified as malware). To validate our results, we also run 10-fold cross-validation, which produced an overall accuracy above 0.997, as shown in Table A.4.

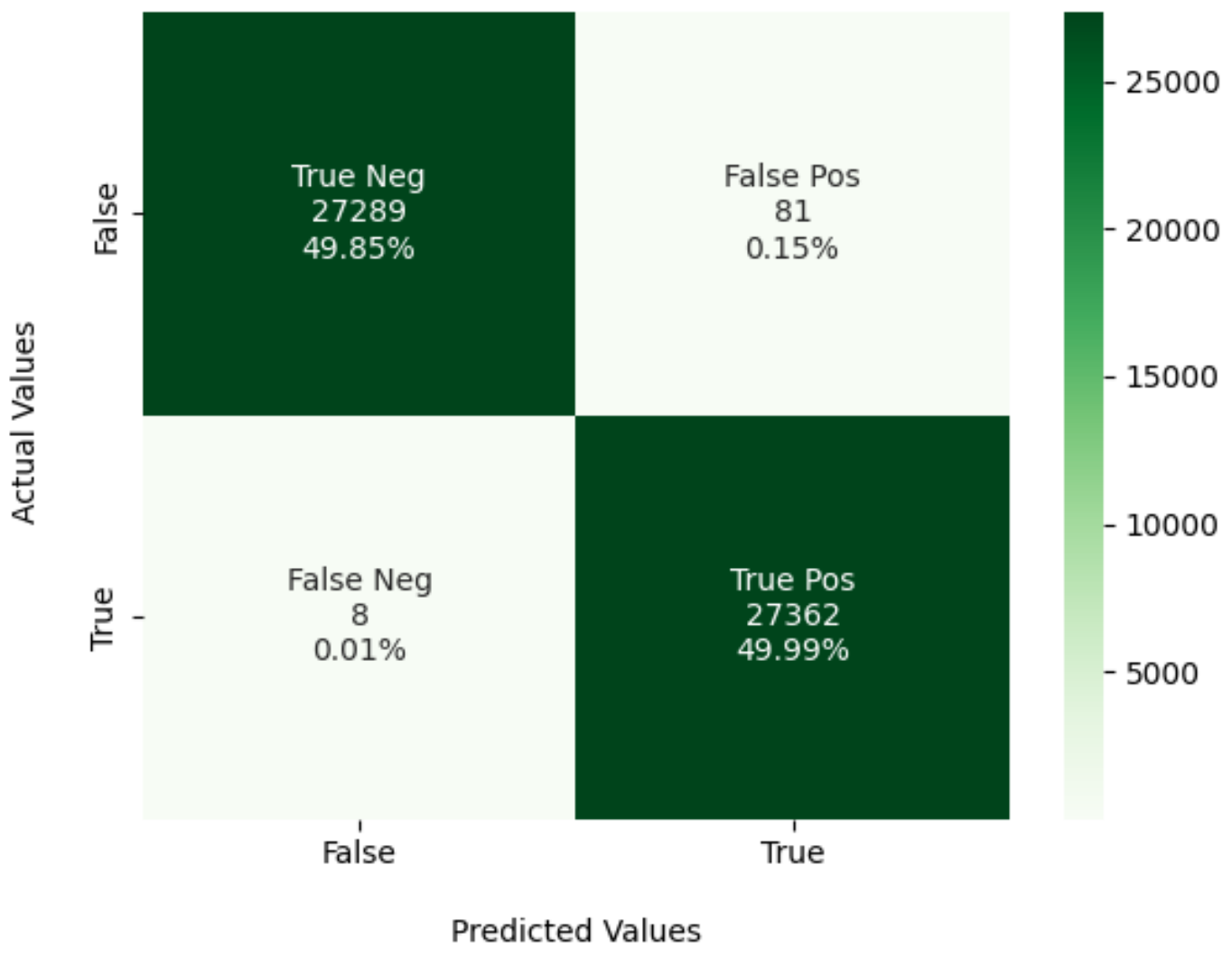

*Figure 3. Confusion matrix for the Random Forest-based malware detection model trained solely on the Transponder malware subtype.*

## 4.3. Model explainability

*Global interpretation*

One of the aims of this study is to ensure that the model's performance is based on interpretable and explainable processes. In this study, the impact of each feature on the prediction of the model was further explored using SHapley Additive exPlanations (SHAP) values, as explained in Section 3.3.3. Here, the SHAP values of each feature are calculated using the test dataset. Since training with the Transponder malware subtype produces the best model performance, we will base model interpretation on its respective model. Figure 4 illustrates a SHAP beeswarm plot of the top five features used by this model and tested on data containing other unseen malware subtypes. In this plot, features are represented vertically along the Y-axis and are ranked in the order of importance as determined by the SHAP analysis. SHAP values for a given dump file and feature are represented horizontally along the X-axis by coloured dots. A red dot represents a high feature value, while a blue dot represents a low feature value. If the value corresponding to a feature is on the left side of the X-axis, it suggests that the feature has a "negative" drag on the prediction of that data instance, thereby pushing the prediction value towards 'benign' classification. On the other hand, if the value corresponding to a feature is on the right side of the X-axis, this means it has a "positive" drag on the prediction, thereby pushing the prediction value towards the ‘malware’ classification. The overall distribution of the SHAP values determines the influence of each feature on the model’s prediction.

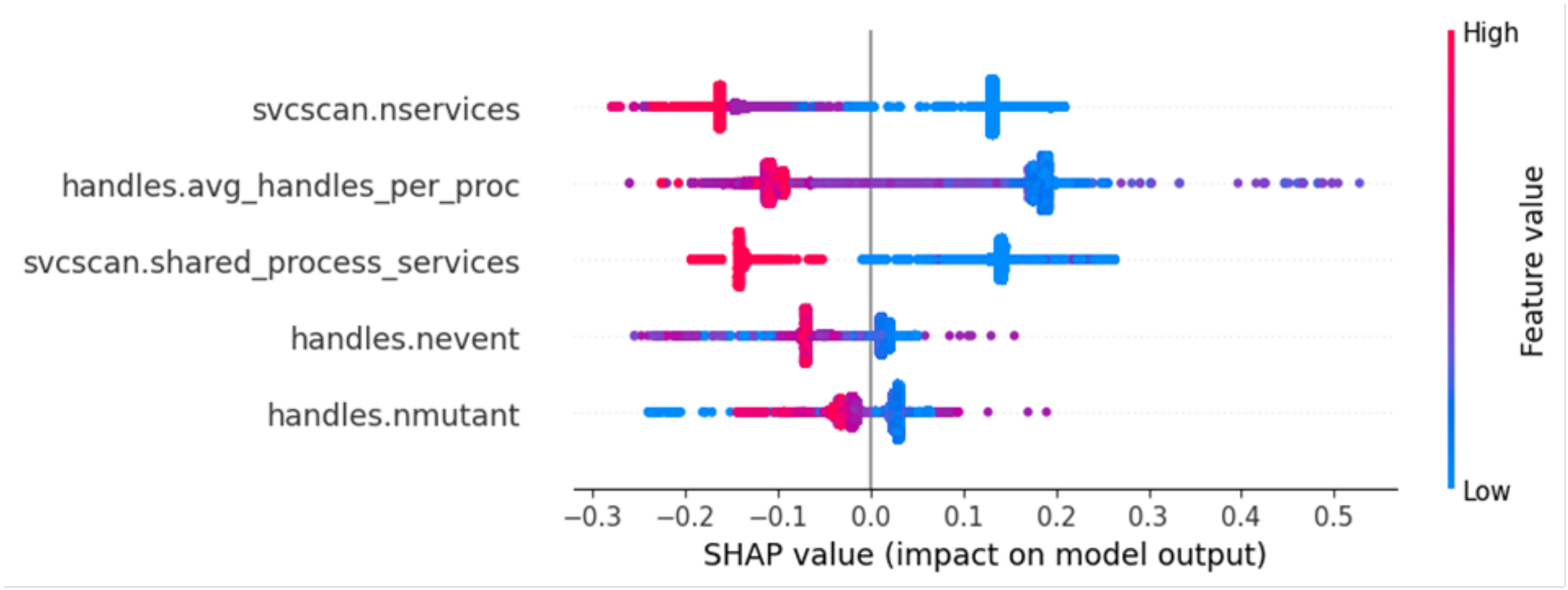

*Figure 4. SHAP Beeswarm plot of the top five features in the Transponder-based malware detection model. Each dot represents a prediction for one file. The position of a dot relative to the X-axis indicates the feature’s influence on that prediction: dots on the left suggest influence towards benign, while dots on the right suggest influence towards malware. The colour of a dot indicates the feature’s value, with red representing a high value and blue a low value.*

Figure 4 indicates that *svcscan.nservices* (total number of services), *handles.avg_handles_per_proc* (average number of handles per process) and *svcscan.shared_process_service* (number of shared processes detected in the memory image) are the topmost critical features for binary classification. Since most blue dots appear on the right side of the X-axis, low numbers of service handles per process and shared processes are associated with a higher likelihood of being classified as malware, in line with behaviour patterns of modern obfuscated malware families. That is, instead of running as independent services or processes, modern obfuscated malware can use techniques such as file-less execution or executing arbitrary code in the address space of a separate live service or process. By suspending an existing service or process, and then unmapping/hollowing its memory, which can then be replaced with malicious code or the path to a DLL. In other words, an obfuscated malware does not need to initiate new services or processes, thereby keeping the number of services or processes low while remaining undetected (Alani et al., 2023; Brennan, 2021).

*Local interpretation*

To determine the contribution of each feature to the prediction for a given instance, the SHAP force plot can be used for local interpretation of each instance. The SHAP force plot offers an in-depth perspective of SHAP values for individual instances and can be used to identify the main features affecting the prediction and the magnitude of its contribution. Figure 5 illustrates the interpretation of SHAP force plot for one benign instance (panel A) and one (Pysa) malware instance (panel B). The model's decision is decomposed into the sum of the effects of each feature value. In the case of the benign instance (Figure 5A), *svcscan.shared_process_service*, *svcscan.nservices* and *handles.avg_handles_per_proc* play pivotal roles, as denoted by the size of the blue arrows, in influencing the model's likelihood of prediction towards benign classification. *handles.nevent* is the feature with the least influence for this memory dump instance. An identical feature influence pattern is observed for the instance featuring Pysa malware (Figure 5B), as indicated by the size of the red arrows.

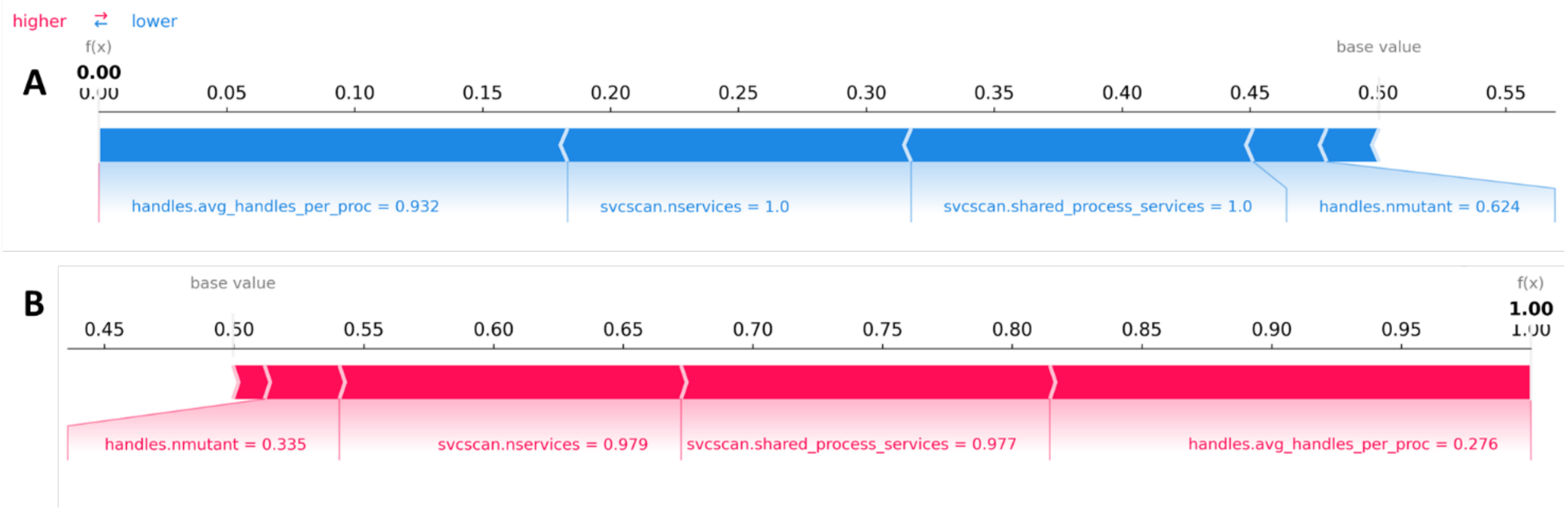

*Figure 5. Local explanations using SHAP force plot for individual prediction cases. Panel A displays a benign instance and Panel B a Pysa malware instance. The f(x) value represents the model's predicted label. The size of each arrow bar for each feature represents the contribution of that feature to the prediction. The colour of the arrow bar indicates the feature's influence: blue suggests influence towards benign, while red suggests influence towards malware.*

# 5. General discussion

## 5.1. Summary of findings

The primary aim of this study was to develop not only a highly accurate, lightweight and interpretable ML-based system for detecting obfuscated malware (see, for example, Alani et al., 2023), but also one that can adapt to previously unknown malware variants. Our novel approach achieved state-of-the-art accuracy despite training on a single malware subtype—namely, the Transponder—and a small subset of features, distinguishing itself in the landscape of malware detection research.

The comparative analysis of our model with existing research is shown in Table 5. Our model demonstrates performance in line with previous work, with accuracy surpassing 99.8%, while trained solely on the Transponder, representing a small fraction of no more than 6.6% of the data. This model was used to successfully classify memory dumps from 14 other unseen obfuscated malware subtypes. This finding indicates that our model is highly adaptive with capacity to generalise beyond training data. Such aspect is crucial in the pursuit of developing defence systems capable of combating zero-day attacks without pre-existing signatures (that is, for detecting emerging vulnerabilities); yet to the best of our knowledge, none of the previous studies based on the CIC-MalMem-2022 dataset, or generally on memory analysis-based features, have considered zero-day attacks when evaluating their ML models.

The size of a malware detection model can present challenges for the implementation on resource-constrained devices. Our model's size of 340 KB is significantly smaller compared to most sizes reported in the literature. For example, Shafin et al. (2023) proposed model sizes of 967 KB for RobustCBL and 575 KB for CompactCBL. Meanwhile, Dener et al. (2022) reported size of 6037 KB for their DNN, while Mezina and Burget (2022) reported a size of 627 KB for their LSTM model.

*Table 5. Comparison of proposed system with related works that used the CIC-MalMem-2022 dataset.*

| Research | Malware Subtype Used in Training | Training Instances | Classifier | Features | Accuracy | Average Detection Speed (µs) | Interp-retable | Zero Day Attacks |
|---|---|---|---|---|---|---|---|---|
| Carrier et al. (2022) | All | 80% (46876) | Ensemble | 55 | 99% | 8 | x | x |
| Dener et al. (2022) | All | 70% (41017) | Logistic Regression | 52 | 99.97% | - | x | x |
| Mezina and Burget (2022) | All | 80% (46876) | Dilated Convolutional Neural Network | 55 | 99.89% | 738 | x | x |
| Shafin et al. (2023) | All | 80% (46876) | RobustCBL<br>CompactCBL | 55 | 99.96%<br>99.92% | 384<br>255 | x | x |
| Roy et al. (2023) | All | 80% (46876) | Hybrid Stack Ensemble | 16 | 99.98% | - | x | x |
| Alani et al. (2023) | All | 66.66% (39060) | Extreme Gradient Boost | 5 | 99.85% | 0.4 | ✓ | x |
| **This Study** | **Transponder only** | **6.58% (3856)** | **Random Forest** | **5** | **99.84%** | **5.7** | ✓ | ✓ |

While emphasising high accuracy and a lightweight model size is crucial, it is equally important for a real-time malware monitoring system to possess the capability to promptly and effectively counter incoming attacks. Our proposed model demonstrated capacity in swiftly identifying individual instances of both existing and unseen obfuscated malware, with an overall average speed of 5.7 µs. This microsecond-level speed surpasses substantially the millisecond-level speeds reported by deep learning-focused studies such as those by Mezina and Burget (2022) (0.738 ms), Shafin et al. (2023) (0.255 ms for CompactCBL and 0.384 ms for RobustCBL). The reported processing speed is comparable to the speeds achieved by ML-based methods like those proposed by Carrier (2021) (8 µs) and Alani et al. (2023) (0.413 µs). The lightweight, accurate, and fast detection speed capabilities of our proposed model suggest the potential for integrating such a model as an effective malware detection system in resource-constrained systems such as IoT devices. To provide an in-depth explanation of the predictions made by our proposed model, the SHAP technique was used to provide both global and local interpretations of the impact of each feature on the model's predictions. The interpretation of the SHAP plot of the top features suggests that low values of features such as *svcscan.nservices*, *handles.avg_handles_per_proc* and *svcscan.shared_process_service* result in the likelihood of prediction towards malware classification, in line with observed behaviour patterns of modern obfuscated-malware families (Brennan, 2021). Local interpretation was also demonstrated; such an approach has the capacity of enabling more targeted and effective security incident responses by providing a detailed understanding of the factors driving a model's prediction for a specific file analysis.

## 5.2. Implications for cyber security defences

Our research addresses a critical gap in cybersecurity by developing a malware detection model that uniquely combines high accuracy, lightweight design and generalisability (Stallings, 2023, pp. 438–448). Traditional high-accuracy models typically require substantial computational resources, limiting their practicality in resource-constrained environments such as IoT devices, mobile systems and embedded security frameworks. On the other hand, lightweight models often compromise on accuracy, reducing their effectiveness against novel or sophisticated malware specimens.

Our model overcomes these limitations by providing robust detection capabilities without a heavy computational footprint, potentially enabling seamless integration into diverse cyber security frameworks. Its lightweight architecture offers an important deployment flexibility advantage, particularly in edge computing and cloud-based applications, where efficiency is crucial for scalability and real-time threat response. Therefore, this approach supports adaptive, resilient and scalable cyber security solutions, aligning with modern industry needs for effective threat mitigation.

Furthermore, the results of this study provide interesting insights for enhancing cyber security defences, particularly within the context of the dynamic and evolving threat landscape. Identification of specific memory processes such as *svcscan.nservices* and *svcscan.shared_process_service* as primary targets across a wide range of malware subtypes, such as Transponder, Zeus and Gator, highlights a potential vulnerability within system architectures (Table A.2). Our findings indicate that these processes, which are critical to operating system service management, are often exploited by attackers aiming to manipulate or disrupt system functions covertly. Developing advanced malware detection systems that specifically guard these service-related processes could significantly mitigate risks posed by such targeted attacks.

The impressive generalisation capability of our model, particularly when trained on the Transponder subtype, to detect other malware variants points to another important advantage of integrating ML in cyber security defences—adaptability. This adaptability feature is a clear differentiating advantage over signature-based or heuristic-based approaches, which require regular updates to address new threats (Al-Asli and Ghaleb, 2019). In our study, we integrated ML only with a memory analysis approach. It remains to be seen if the integration of ML with static analysis could lead to the same adaptability advantage and could offer insights into malware's structural features without execution (Aslan and Samet, 2020). Such pursuit with all malware analysis approaches could result in a comprehensive detection framework that maximises the strengths of each analysis approach, fostering robust defence mechanisms against diverse malware types.

Finally, the finding that some malware subtypes, particularly Transponder, exhibit superior generalisation for detecting unseen malware suggests that intrinsic characteristics of these malware—such as its operational framework or attack strategies—might play a crucial role in providing informative training data for the extraction of malware attack patterns generalisable across different malware variants (at least when memory-analysis is used). Our work highlights the need for cyber security experts to investigate why Transponder generalises better than other subtypes like Zeus, or at a broader scale, why the Spyware is more effective at detecting unseen malware compared to other types, such as Trojan horse or Ransomware (see Figure 2).

## 5.3. Limitations and future directions

While our study presents a robust proof-of-concept for an adaptive malware detection system, several limitations could be addressed to improve the system's effectiveness. Firstly, we did not perform hyperparameter tuning, instead opting for default hyperparameter settings for our classifiers. Our primary objective was to demonstrate the potential of ML-based approaches for detecting unknown malware attacks. However, further refinement through hyperparameter tuning and testing of alternative feature selection methods could yield improved performance and generalisation capabilities. We also expect that other ML models, such as Gradient

Boosting Machines, could offer substantial improvements in speed while maintaining similar accuracy levels (see for example, Alani et al., 2023).

To improve model explainability, future work could incorporate error analysis to evaluate when and why the model makes errors in specific cases (Nushi, 2021). This approach would provide insights into the model's weaknesses and guide improvements. Moreover, future research could focus on developing similar models on other obfuscated or non-obfuscated malware datasets. This would test the generalisability of our approach across different types of malware and identify which types of attacks are more amenable to zero-shot learning (Palatucci et al., 2009; i.e., enabling the trained model to correctly make predictions on data from unseen malware subtypes). Finally, training on one dataset and testing on a different dataset could provide further validation of our model's adaptability and robustness.

## 5.4. Conclusion

In this paper, we have presented a novel machine learning-based system for detecting obfuscated malware. Trained on a small dataset from a single malware subtype—Transponder—our system achieves state-of-the-art accuracy, while maintaining rapid processing speeds (5.7 µs per file) and minimal memory usage (a model size of 340 KB). These findings not only advance the state-of-the-art in malware detection, but also highlight the critical need for the testing and refinement of machine learning-based solutions for detecting previously unseen malware variants. This is crucial for ensuring that our defence solutions can respond to the continuously evolving landscape of cyber threats.

## CRediT authorship contribution statement

**Oladipo A. Madamidola**: Methodology, Software, Data curation, Investigation, Writing - Original draft preparation, Visualisation. **Felix Ngobigha**: Conceptualisation, Writing - Review & Editing, Supervision. **Adnane Ez-zizi**: Conceptualisation, Methodology, Validation, Writing - Review & Editing, Visualisation, Supervision.

## Declaration of competing interest

The authors declare that they have no known competing financial interests or personal relationships that could have appeared to influence the work reported in this paper.

## Acknowledgements

We thank Shahroz Nadeem and Muhammad Waqar for comments on earlier drafts of this paper. We are also grateful to the participants of the Technology Seminar at the University of Suffolk for their discussions and feedback during a presentation of this work, conducted by Adnane Ez-zizi.

## Data and code availability

All data and code are available on GitHub at
https://github.com/Adnane017/Detecting_new_obfuscated_malware_variants

## Declaration of generative AI and AI-assisted technologies in the writing process

During the preparation of this work the authors used OpenAI ChatGPT (version 4.0, 2024) in order to proofread and improve the readability of the manuscript. After using this tool, the authors reviewed and edited the content as needed and take full responsibility for the content of the publication.

## References

Acuto, A., Maskell, S. and Jack, D., 2023. Defending the unknown: Exploring reinforcement learning agents' deployment in realistic, unseen networks. In 2023 Conference on Applied Machine Learning in Information Security (pp. 22-35).

Alam, S., Horspool, R.N., Traore, I. and Sogukpinar, I., 2015. A framework for metamorphic malware analysis and real-time detection. *Computers & Security*, *48*, pp.212-233.

Alani, M.M., Mashatan, A. and Miri, A., 2023. XMal: A lightweight memory-based explainable obfuscated-malware detector. *Computers & Security*, *133*, p.103409.

Al-Asli, M. and Ghaleb, T.A., 2019, April. Review of signature-based techniques in antivirus products. In *2019 International Conference on Computer and Information Sciences (ICCIS)* (pp. 1-6). IEEE.

Alazab, M., Venkatraman, S., Watters, P.A. and Alazab, M., 2011. Zero-day Malware Detection based on Supervised Learning Algorithms of API call Signatures. *AusDM*, 11, pp.171-182.

Alduailij, M., Khan, Q.W., Tahir, M., Sardaraz, M., Alduailij, M. and Malik, F., 2022. Machine-learning-based DDoS attack detection using mutual information and random forest feature importance method. Symmetry, 14(6), pp.1–15.

Alhaidari, F., Shaib, N. A., Alsafi, M., Alharbi, H., Alawami, M., Aljindan, R., Rahman, A.-U. and Zagrouba, R., 2022. Zevigilante: Detecting zero-day malware using machine learning and sandboxing analysis techniques*. Computational Intelligence and Neuroscience*, 2022, 1615528.

Amer, E. and Zelinka, I., 2020. A dynamic Windows malware detection and prediction method based on contextual understanding of API call sequence. *Computers & Security*, 92, 101760.

Aslan, Ö.A. and Samet, R., 2020. A comprehensive review on malware detection approaches. *IEEE access*, *8*, pp.6249-6271.

Bhatia, T. and Kaushal, R., 2017, June. Malware detection in android based on dynamic analysis. In *2017 International conference on cyber security and protection of digital services (Cyber security)* (pp. 1-6). IEEE.

Brennan, M. 2021, Cobalt Strikes Again: An Analysis of Obfuscated Malware, blog, Huntress, viewed 22 May 2024, https://www.huntress.com/blog/cobalt-strike-analysis-of-obfuscated-malware.

Carrier, T., Victor, P., Tekeoglu, A. and Lashkari, A.H., 2022, February. Detecting Obfuscated Malware using Memory Feature Engineering. In *Icissp* (pp. 177-188).

Dada, E.G., Bassi, J.S., Hurcha, Y.J. and Alkali, A.H., 2019. Performance evaluation of machine learning algorithms for detection and prevention of malware attacks. *IOSR Journal of Computer Engineering*, *21*(3), pp.18-27.

Degenhardt, F., Seifert, S. and Szymczak, S., 2019. Evaluation of variable selection methods for random forests and omics data sets. *Briefings in Bioinformatics*, *20*(2), pp.492-503.

Dener, M., Ok, G. and Orman, A., 2022. Malware detection using memory analysis data in big data environment. *Applied Sciences*, *12*(17), p.8604.

Elkhail, A.A., Refat, R.U.D., Habre, R., Hafeez, A., Bacha, A. and Malik, H., 2021. Vehicle security: A survey of security issues and vulnerabilities, malware attacks and defenses. *IEEE Access*, *9*, pp.162401-162437.

Gandotra, E., Bansal, D. and Sofat, S., 2016, December. Zero-day malware detection. In *2016 Sixth international symposium on embedded computing and system design (ISED)* (pp. 171-175). IEEE.

Genuer, R., Poggi, J.M. and Tuleau-Malot, C., 2010. Variable selection using random forests. *Pattern recognition letters*, *31*(14), pp.2225-2236.

Guo, Y., 2023. A review of Machine Learning-based zero-day attack detection: Challenges and future directions. *Computer communications*, *198*, pp.175-185.

Gupta, D. and Rani, R., 2020. Improving malware detection using big data and ensemble learning. *Computers & Electrical Engineering*, *86*, p.106729.

Hadiprakoso, R.B., Kabetta, H. and Buana, I.K.S., 2020, November. Hybrid-based malware analysis for effective and efficiency android malware detection. In *2020 International Conference on Informatics, Multimedia, Cyber and Information System (ICIMCIS)* (pp. 8-12). IEEE.

Huang, C.Y., Tsai, Y.T. and Hsu, C.H., 2013. Performance evaluation on permission-based detection for android malware. In *Advances in Intelligent Systems and Applications-Volume 2: Proceedings of the International Computer Symposium ICS 2012 Held at Hualien, Taiwan, December 12–14, 2012* (pp. 111-120). Springer Berlin Heidelberg.

Huda, S., Miah, S., Hassan, M. M., Islam, R., Yearwood, J., Alrubaian, M. and Almogren, A., 2017. Defending unknown attacks on cyber-physical systems by semi-supervised approach and available unlabeled data. *Information Sciences*, 379, 211-228.

Hwang, J., Kim, J., Lee, S. and Kim, K., 2020. Two-stage ransomware detection using dynamic analysis and machine learning techniques. *Wireless Personal Communications*, *112*(4), pp.2597-2609.

Idika, N. and Mathur, A.P., 2007. A survey of malware detection techniques. *Technical Report. Purdue University*. Available at https://www.cerias.purdue.edu/apps/reports_and_papers/view/4328/.

Ijaz, M., Durad, M.H. and Ismail, M., 2019, January. Static and dynamic malware analysis using machine learning. In *2019 16th International Bhurban Conference on Applied Sciences and Technology (IBCAST)* (pp. 687-691). IEEE.

Jain, A. and Singh, A.K., 2017, August. Integrated Malware analysis using machine learning. In *2017 2nd International Conference on Telecommunication and Networks (TEL-NET)* (pp. 1-8). IEEE.

Kim, J.Y., Bu, S.J. and Cho, S.B., 2018. Zero-day malware detection using transferred generative adversarial networks based on deep autoencoders. *Information Sciences*, *460*, pp.83-102.

Lashkari, A.H., Li, B., Carrier, T.L. and Kaur, G., 2021, May. Volmemlyzer: Volatile memory analyzer for malware classification using feature engineering. In *2021 Reconciling Data Analytics, Automation, Privacy, and Security: A Big Data Challenge (RDAAPS)* (pp. 1-8). IEEE.

Liu, X., Lin, Y., Li, H. and Zhang, J., 2020. A novel method for malware detection on ML-based visualization technique. *Computers & Security*, *89*, p.101682.

Lundberg, S.M. and Lee, S.I., 2017. A unified approach to interpreting model predictions. *Advances in Neural Information Processing Systems*, *30*.

Lundberg, S.M., Erion, G., Chen, H., DeGrave, A., Prutkin, J.M., Nair, B., Katz, R., Himmelfarb, J., Bansal, N. and Lee, S.I., 2019. Explainable AI for trees: From local explanations to global understanding. *arXiv preprint arXiv:1905.04610*.

Lundberg, S.M., Erion, G.G. and Lee, S.I., 2018. Consistent individualized feature attribution for tree ensembles. *arXiv preprint arXiv:1802.03888*.

Mezina, A. and Burget, R., 2022, October. Obfuscated malware detection using dilated convolutional network. In *2022 14th international congress on ultra modern telecommunications and control systems and workshops (ICUMT)* (pp. 110-115). IEEE.

Mijwil, M., Salem, I.E. and Ismaeel, M.M., 2023. The significance of machine learning and deep learning techniques in cybersecurity: A comprehensive review. *Iraqi Journal For Computer Science and Mathematics*, *4*(1), pp.87-101.

Mirsky, Y., Doitshman, T., Elovici, Y. and Shabtai, A., 2018. Kitsune: an ensemble of autoencoders for online network intrusion detection. *arXiv preprint arXiv:1802.09089*.

Mosli, R., Li, R., Yuan, B. and Pan, Y., 2016, May. Automated malware detection using artifacts in forensic memory images. In *2016 IEEE Symposium on Technologies for Homeland Security (HST)* (pp. 1-6). IEEE.

Niu, D., Wang, K., Sun, L., Wu, J. and Xu, X., 2020. Short-term photovoltaic power generation forecasting based on random forest feature selection and CEEMD: A case study. *Applied soft computing*, *93*, p.106389.

Nushi, B. 2021, Responsible Machine Learning with Error Analysis, blog, Microsoft Tech Community, viewed 16 May 2024, https://techcommunity.microsoft.com/t5/ai-machine-learning-blog/responsible-machine-learning-with-error-analysis/ba-p/2141774.

O'Kane, P., Sezer, S. and McLaughlin, K., 2011. Obfuscation: The hidden malware. *IEEE Security & Privacy*, *9*(5), pp.41-47.

Palatucci, M., Pomerleau, D., Hinton, G.E. and Mitchell, T.M., 2009. Zero-shot learning with semantic output codes. Advances in neural information processing systems, 22.

Rad, B.B., Masrom, M. and Ibrahim, S., 2012. Camouflage in malware: from encryption to metamorphism. *International Journal of Computer Science and Network Security*, *12*(8), pp.74-83.

Radwan, A.M., 2019, October. Machine learning techniques to detect maliciousness of portable executable files. In *2019 International Conference on Promising Electronic Technologies (ICPET)* (pp. 86-90). IEEE.

Rathnayaka, C. and Jamdagni, A., 2017, August. An efficient approach for advanced malware analysis using memory forensic technique. In *2017 IEEE Trustcom/BigDataSE/ICESS* (pp. 1145-1150). IEEE.

Roy, K.S., Ahmed, T., Udas, P.B., Karim, M.E. and Majumdar, S., 2023. Malhystack: A hybrid stacked ensemble learning framework with feature engineering schemes for obfuscated malware analysis. *Intelligent Systems with Applications*, *20*, p.200283.

Shafin, S.S., Karmakar, G. and Mareels, I., 2023. Obfuscated memory malware detection in resource-constrained IoT devices for smart city applications. *Sensors*, *23*(11), p.5348.

Sihwail, R., Omar, K., Zainol Ariffin, K.A. and Al Afghani, S., 2019. Malware detection approach based on artifacts in memory image and dynamic analysis. *Applied Sciences*, *9*(18), p.3680.

Sihwail, R., Omar, K. and Arifin, K.A.Z., 2021. An Effective Memory Analysis for Malware Detection and Classification. *Computers, Materials & Continua*, *67*(2).

Speiser, J.L., Miller, M.E., Tooze, J. and Ip, E., 2019. A comparison of random forest variable selection methods for classification prediction modeling. *Expert systems with applications*, *134*, pp.93-101.

Stallings, W. (2023). Cryptography and network security: principles and practice (8th ed., pp. 438–448). Harlow, United Kingdom: Pearson.

Tahir, R., 2018. A study on malware and malware detection techniques. *International Journal of Education and Management Engineering*, *8*(2), p.20.

Wang, T.Y., Horng, S.J., Su, M.Y., Wu, C.H., Wang, P.C. and Su, W.Z., 2006, July. A surveillance spyware detection system based on data mining methods. In *2006 IEEE International Conference on Evolutionary Computation* (pp. 3236-3241). IEEE.

Zhao, Y., Zhu, W., Wei, P., Fang, P., Zhang, X., Yan, N., Liu, W., Zhao, H. and Wu, Q., 2022. Classification of Zambian grasslands using random forest feature importance selection during the optimal phenological period. *Ecological Indicators*, *135*, p.108529.

Zhou, Q. and Pezaros, D., 2019. Evaluation of Machine Learning Classifiers for Zero-Day Intrusion Detection–An Analysis on CIC-AWS-2018 dataset. *arXiv preprint arXiv:1905.03685*.

# Appendix

*Table A.1. List of features in the CIC-Malmem-2022 dataset.*

| Feature Category | Feature | Description |
|---|---|---|
| Callbacks | callbacks.**nanonymous** | Total number of anonymous callbacks. |
| | callbacks.**ncallbacks** | Total number of registered callback functions. |
| | callbacks.**ngeneric** | Total number of generic callbacks. |
| DLLlist | dlllist.**avg_dlls_per_proc** | Average number of DLLs loaded per process. |
| | dlllist.**ndlls** | Total number of loaded DLLs. |
| Handles | handles.**avg_handles_per_proc** | Average number of handles per process. |
| | handles.**ndesktop** | Total number of desktop handles. |
| | handles.**ndirectory** | Total number of directory handles. |
| | handles.**nevent** | Total number of event handles. |
| | handles.**nfile** | Total number of file handles. |
| | handles.**nhandles** | Total number of handles opened. |
| | handles.**nkey** | Total number of registry key handles. |
| | handles.**nmutant** | Total number of mutant handles. |
| | handles.**nport** | Total number of port handles. |
| | handles.**nsection** | Total number of section handles. |
| | handles.**nsemaphore** | Total number of semaphore handles. |
| | handles.**nthread** | Total number of thread handles. |
| | handles.**ntimer** | Total number of timer handles. |
| LDR modules | ldrmodules.**not_in_init** | Total number of modules not in the initialised state. |
| | ldrmodules.**not_in_init_avg** | Average number of modules not initialised. |
| | ldrmodules.**not_in_load** | Total number of modules not in a loaded state. |
| | ldrmodules.**not_in_load_avg** | Average number of modules not loaded. |
| | ldrmodules.**not_in_mem** | Total number of modules not present in memory. |
| | ldrmodules.**not_in_mem_avg** | Average number of modules not in memory. |
| MalFind | malfind.**commitCharge** | Total number of commit charge of detected memory injections. |
| | malfind.**ninjections** | Total number of memory injections detected. |
| | malfind.**protection** | Total number of protection attributes of detected memory injections. |
| | malfind.**uniqueInjections** | Total number of unique memory injections detected. |
| Modules | modules.**nmodules** | Total number of loaded modules. |
| PsList | pslist.**avg_handlers** | Average number of handlers per process. |
| | pslist.**avg_threads** | Average number of threads per process. |
| | pslist.**nppid** | Total number of parent processes. |
| | pslist.**nproc** | Total number of processes running. |
| | pslist.**nprocs64bit** | Total number of 64-bit processes running. |
| Psxview | psxview.**not_in_csrss_handles** | Total number of processes not in CSRSS handles. |
| | psxview.**not_in_csrss_handles_false_avg** | Average number of processes not in CSRSS handles. |
| | psxview.**not_in_deskthrd** | Total number of processes not in the desktop thread. |
| | psxview.**not_in_deskthrd_false_avg** | Average number of processes not in the desktop thread. |
| | psxview.**not_in_eprocess_pool** | Total number of processes not in the EPROCESS pool. |
| | psxview.**not_in_eprocess_pool_false_avg** | Average number of processes not in the EPROCESS pool. |
| | psxview.**not_in_ethread_pool** | Total number of processes not in the ETHREAD pool. |
| | psxview.**not_in_ethread_pool_false_avg** | Average number of processes not in the ETHREAD pool. |
| | psxview.**not_in_pslist** | Total number of processes not in the process list. |
| | psxview.**not_in_pslist_false_avg** | Average number of processes not in the process list. |
| | psxview.**not_in_pspcid_list** | Total number of processes not in the PSPCID list. |
| | psxview.**not_in_pspcid_list_false_avg** | Average count of processes not in the PSPCID list. |
| | psxview.**not_in_session** | Total number of processes not in any session. |
| | psxview.**not_in_session_false_avg** | Average number of processes not in any session. |
| SVCScan | svcscan.**fs_drivers** | Total number of file system drivers. |
| | svcscan.**interactive_process_services** | Total number of services in interactive processes. |
| | svcscan.**kernel_drivers** | Total number of kernel drivers. |
| | svcscan.**nactive** | Total number of services running. |
| | svcscan.**nservices** | Total number of services running. |
| | svcscan.**process_services** | Total number of services in separate processes. |
| | svcscan.**shared_process_services** | Total number of services in shared processes. |

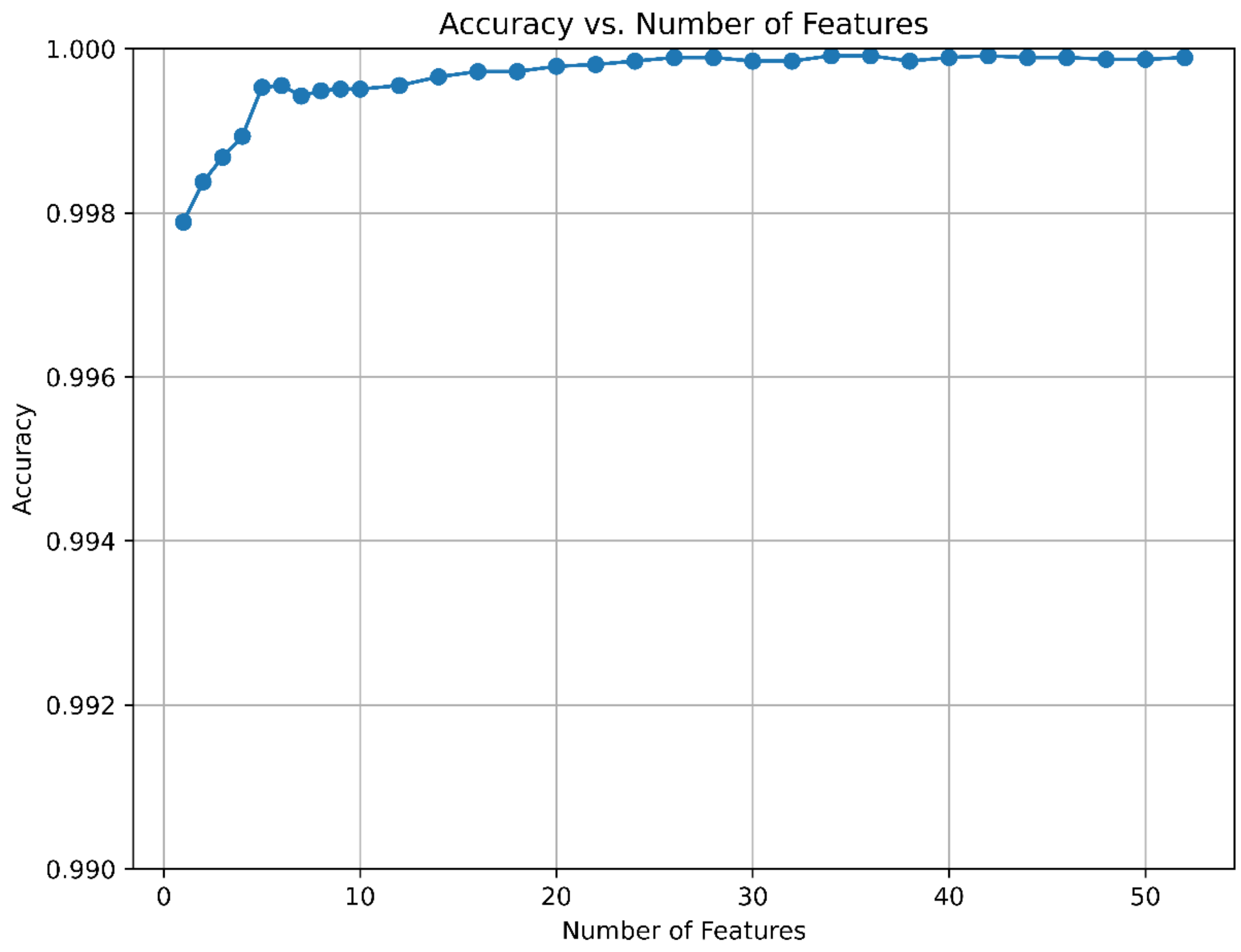

*Figure A.1. Change in Accuracy with Number of Features.*

*Table A.2. Results of 10-fold Cross-Validation for Random Forest.*

| Fold | Accuracy | Precision | Recall | F1-Score |
|---|---|---|---|---|
| **1** | 0.999829 | 0.999829 | 0.999829 | 0.999829 |
| **2** | 1.000000 | 1.000000 | 1.000000 | 1.000000 |
| **3** | 0.999488 | 0.999488 | 0.999488 | 0.999488 |
| **4** | 1.000000 | 1.000000 | 1.000000 | 1.000000 |
| **5** | 0.999829 | 0.999829 | 0.999829 | 0.999829 |
| **6** | 1.000000 | 1.000000 | 1.000000 | 1.000000 |
| **7** | 1.000000 | 1.000000 | 1.000000 | 1.000000 |
| **8** | 1.000000 | 1.000000 | 1.000000 | 1.000000 |
| **9** | 1.000000 | 1.000000 | 1.000000 | 1.000000 |
| **10** | 1.000000 | 1.000000 | 1.000000 | 1.000000 |
| **μ** | 0.999915 | 0.999915 | 0.999915 | 0.999915 |
| **σ** | 0.000166 | 0.000166 | 0.000166 | 0.000166 |

*Table A.3. Summary of results from training on each malware subtype. For each trained model, the table includes the average accuracy, model size in memory and the top five features in order of importance. Additionally, the ranking of the model relative to the other subtype-focused models is indicated between parentheses.*

| | **Zeus (15)** | **Emotet (9)** | **Refroso (13)** | **Scar (14)** | **Reconyc (5)** |
|---|---|---|---|---|---|
| **Trojan Horse** | **1.** pslist.avg_handlers<br>**2.** handles. avg_handles_per_proc<br>**3.** svcscan.nservices<br>**4.** svcscan. shared_process_services<br>**5.** handles.nhandles | **1.** svcscan.nservices<br>**2.** svcscan. shared_process_services<br>**3.** svcscan.kernel_drivers<br>**4.** slist.avg_handlers<br>**5.** handles.nsection | **1.** handles. avg_handles_per_proc<br>**2.** pslist.avg_handlers<br>**3.** svcscan.nservices<br>**4.** svcscan. shared_process_services<br>**5.** dlllist.avg_dlls_per_proc | **1.** pslist.avg_handlers<br>**2.** handles. avg_handles_per_proc<br>**3.** svcscan.nservices<br>**4.** svcscan. shared_process_services<br>**5.** handles.nhandles | **1.** svcscan.nservices<br>**2.** svcscan. shared_process_services<br>**3.** handles.nsection<br>**4.** handles.nmutant<br>**5.** svcscan.kernel_drivers |
| **Accuracy Model Size** | **0.9800**<br>90.21 KB | **0.9901**<br>134.74 KB | **0.9810**<br>97.26 KB | **0.9810**<br>119.43 KB | **0.9967**<br>169.59 KB |
| | **180Solutions (4)** | **Coolwebsearch (10)** | **Gator (2)** | **Transponder (1)** | **TIBS (7)** |
| **Spyware** | **1.** svcscan.nservices<br>**2.** handles.nevent<br>**3.** svcscan. shared_process_services<br>**4.** handles.nmutant<br>**5.** svcscan.kernel_drivers | **1.** svcscan.nservices<br>**2.** pslist.avg_handlers<br>**3.** svcscan. shared_process_services<br>**4.** svcscan.kernel_drivers<br>**5.** handles. avg_handles_per_proc | **1.** svcscan.nservices<br>**2.** svcscan. shared_process_services<br>**3.** handles. avg_handles_per_proc<br>**4.** pslist.avg_handlers<br>**5.** handles.nmutant | **1.** svcscan.nservices<br>**2.** handles. avg_handles_per_proc<br>**3.** svcscan. shared_process_services<br>**4.** handles.nevent<br>**5.** handles.nmutant | **1.** svcscan.nservices<br>**2.** svcscan. shared_process_services<br>**3.** handles.nhandles<br>**4.** handles.nevent<br>**5.** handles.nmutant |
| **Accuracy Model Size** | **0.9972**<br>154.12 KB | **0.9904**<br>129.29 KB | **0.9977**<br>148.49 KB | **0.9984**<br>339.59 KB | **0.9955**<br>124.42 KB |

| | Conti (6) | MAZE (8) | Pysa (3) | Ako (12) | Shade (11) |
|---|---|---|---|---|---|
| **Ransomware** | **1.** svcscan.kernel_drivers<br>**2.** svcscan.nservices<br>**3.** svcscan. shared_process_services<br>**4.** handles.nsection<br>**5.** handles.nevent | **1.** handles.nhandles<br>**2.** svcscan. shared_process_services<br>**3.** svcscan.nservices<br>**4.** svcscan.kernel_drivers<br>**5.** handles.nmutant | **1.** svcscan.nservices<br>**2.** svcscan. shared_process_services<br>**3.** pslist.avg_handlers<br>**4.** svcscan.kernel_drivers<br>**5.** handles.nevent | **1.** svcscan.nservices<br>**2.** pslist.avg_handlers<br>**3.** svcscan.shared_process_services<br>**4.** dlllist.avg_dlls_per_proc<br>**5.** svcscan.kernel_drivers | **1.** svcscan.nservices<br>**2.** svcscan. shared_process_services<br>**3.** svcscan.kernel_drivers<br>**4.** handles.nevent<br>**5.** dlllist.avg_dlls_per_proc |
| **Accuracy Model Size** | **0.9962**<br>150.68 KB | **0.9953**<br>147.09 KB | **0.9974**<br>128.81 KB | **0.9844**<br>113.02 KB | **0.9905**<br>116.46 KB |

Table A.4. Results of 10-fold Cross-Validation for Random Forest-based malware detection model trained solely on the Transponder malware subtype.

| Fold | Accuracy | Precision | Recall | F1-Score |
|---|---|---|---|---|
| **1** | 1.000000 | 1.000000 | 1.000000 | 1.000000 |
| **2** | 0.997409 | 0.997423 | 0.997409 | 0.997409 |
| **3** | 0.992228 | 0.992241 | 0.992228 | 0.992228 |
| **4** | 0.997409 | 0.997423 | 0.997409 | 0.997409 |
| **5** | 0.994819 | 0.994872 | 0.994819 | 0.994819 |
| **6** | 0.994819 | 0.994819 | 0.994819 | 0.994819 |
| **7** | 0.997403 | 0.997416 | 0.997403 | 0.997403 |
| **8** | 0.997403 | 0.997416 | 0.997403 | 0.997403 |
| **9** | 1.000000 | 1.000000 | 1.000000 | 1.000000 |
| **10** | 0.997403 | 0.997416 | 0.997403 | 0.997403 |
| **μ** | 0.996889 | 0.996903 | 0.996889 | 0.996889 |
| **σ** | 0.002380 | 0.002374 | 0.002380 | 0.002380 |